\documentclass[11point]{article}
\begin{document}
\title{\Large  \textbf{ Stability and Genericity Aspects of Properties of Space-times in General Relativity}}
\date{}
\maketitle{
\begin{center}
R.V. Saraykar\\
Department of Mathematics, RTM Nagpur University, University Campus, Nagpur-440033, India\\
Mail : ravindra.saraykar$@$gmail.com\vspace{0.5cm}\\
\end{center}
\vspace{10mm}

\noindent\textbf{Abstract}: In this article, we review and discuss different aspects of stability and genericity of some properties of  space-times
which occur in various contexts in the General Theory of Relativity. We also give an argument supporting the conclusion that `Linearization Stability' is a generic property if we restrict space-times to the class of those which admit compact spacelike constant mean curvature hypersurfaces.\\

\section{Introduction}

The famous Japanese astronomer Yusuke Hagihara who is remembered for his contributions to celestial mechanics, formulated the  problem of stability of solar system in the following manner [1]  : He asked ,"Will the present configuration of the solar system be preserved for some long interval of time ? Will the planets eventually fall into the Sun or will some of the planets recede gradually from the Sun so that they no longer belong to the solar system ? What is the interval of time at the end of which the solar system deviates from the present configuration by a previously assigned small amount ?  "He further makes significant remarks"  : "The question has long been an acute problem in celestial mechanics since Laplace. The term 'stability' has often been discussed by various mathematicians and the solution of the problem becomes more and more complicated and difficult to answer as we dig deeper and deeper into it. Present day mathematics hardly enables us to answer this question in a satisfactory manner for the actual solar system. We must limit ourselves here to describe the present status of the efforts towards solving this fascinating but difficult problem of human culture."
More than fifty years after Hagihara's remarks as above, situation remains more or less the same, though  this quest led to many mathematical developments, and several successive 'proofs' of stability for the solar system.\\
More importantly, the problem of stability of solar system forms a small subset of stability research. Prigogine [2] pointed out that the origin of life may be related to successive instabilities and bifurcations. In thermodynamics and statistical mechanics we study the role of entropy and in modern theory of dynamics, we describe chaotic motion and instability by Kolmogorov's entropy.\\
Stability issues play an important role not only in celestial mechanics, but also in fluid dynamics, economical models, numerical algorithms, general theory of relativity, quantum mechanics, nuclear physics, dynamical systems, and control theory as successfully applied in the fields of mechanical and electrical engineering. In the article by V. Szebehely [3], the author discussed more general ideas regarding stability. He remarks "Consider the stability of theorems in any field of science. Often theorems are formulated as follows : 'If certain conditions are given, then a specific statement is true.' We might ask by whom are these conditions given and how well are they fulfilled in any actual application ? Small errors and uncertainties in the 'given conditions' might destroy the validity of the conclusion if the theorem is unstable to small disturbances. Such theorem-stability might be applicable to the establishment of physical laws which are often based on observations with uncertainties. What errors are allowed and how do we estimate the actual errors and uncertainties in order not to destroy the validity of the stated law as the consequence of observations? Clearly, the sensitivity of laws and theorems to observational errors or to the uncertainties of the conditions is critical."\\
Szebehely further observes, " One might go one step further and state that depending on the specific definition of stability used, real systems may always present instabilities when suitably large disturbances are introduced. Once again, the important idea is to find the "proper" disturbance and the "proper" stability condition when a given system or phenomena is investigated. We must realize that stability depends on how much disturbance or deviation is allowed when the system is still called "stable". Furthermore, disturbances should be of relevant type and physical repeatability should be demanded only for striking features. This leads us to the idea of qualitative versus quantitative stability."\\
The development of concepts of stability and stability theorems in various fields developed over a period of past four decades prove that the comments by Szebehely are really true. Moreover, we must note that Szebehely himself, in his article, has given fifty odd terminologies and concepts of stability which are regularly used in the literature. He also suggests that they all are not unique, and in fact are often contradictory and / or repetitive. However, the fact remains that original concepts of stability were given by Lagrange, Poisson, Poincare and Lyapunov. The most well known definition of stability is that of Lyapunov which is the stability of a mechanical system ( system of ordinary or partial differential equations ) with respect to initial conditions.
This definition stood for more than 100 years since the time Lyapunov introduced it in 1907, with modifications over a period of time, especially after 1930. The review article by Loria and Penteley [4] describes in details the development of the concept of stability from historical perspective. Very interestingly, they remark : "We insist on the fact that only the attentive reading of the original documents can contribute to correct certain errors endlessly repeated by different authors."\\
For a more recent review about historical development of classical stability concepts, we refer the reader to R.I. Leine [5].\\
In the present article, we wish to discuss some of the stability issues which have been studied by various researchers in the general theory of relativity. S.W. Hawking [6] has emphasized the reason for studying stable properties as follows : A physical theory is a correspondence between certain physical observations and a mathematical model. In this case, the model is a manifold with a Lorentz metric. The accuracy of the observations is always limited by practical difficulties and by the uncertainty principle. Thus the only properties of space-time which are physically significant are those that are stable in an appropriate topology. Unstable properties will not have physical relevance. Moreover, they may be of mathematical inconvenience in the sense that they may provide counter examples to general theorems one would like to prove about all metrics in a certain region of $Lor(M)$. Here $Lor(M)$ denotes the set of all Lorentz metrics which can be defined on a given space-time manifold. This means the theorem may hold for almost all metrics in the region, but fail for some particular metrics. In this situation, we say that such a theorem holds generically. In general, a property is generic in a region of $Lor(M)$ if it holds almost everywhere on that region. Mathematically, by 'almost everywhere' we mean that it holds on an open dense subset of the region of $Lor(M)$. For physical purposes, it is sufficient to prove that a theorem holds generically because the metric of mathematical model of space-time is defined only with limited accuracy. Hawking [6] defines stability of a property in the language of open subsets of the set $Lor(M)$ under suitable topology. Similar definition of stability is in use in the theory of dynamical systems also.\\
From the definitions of "Stability" and "genericity", it is clear that these concepts depend upon the topologies under consideration. Thus a given property may be stable and generic in some topologies and not in others. Which of the topologies is of physical interest will depend upon the nature of the property under consideration.\\
Thus, in this article, we study and discuss (i) the stability of global properties of space-time, especially causal properties, and (ii) the stability of the outcome of gravitational collapse in spherically symmetric space-times with respect to small perturbations in the regular initial data from which the collapse evolves. Recent article by Fletcher [7] throws more light on the choice of topology and hence on the results about stable and generic properties in this context. Earlier work by other researchers is also discussed. In the later case, we mainly discuss stability of outcome of inhomogeneous dust collapse with respect to initial conditions by referring to the work of Saraykar and Ghate [8] and Joshi, Malafarina and Saraykar [9,10]. \\ 
Furthermore, we discuss the issue of genericity which occurs in general theory of relativity in different contexts. Genericity issue plays an equally important role as stability in Mathematics, Mathematical and Physical sciences, and other sciences also.\\ 
When we define a certain property of a topological space, or a measure space, or property of a dynamical system or that of a space-time in the theory of relativity, we expect that such a property will hold for "almost all" objects.  The word "generic" is used to describe such properties. Thus, properties that hold for "typical" examples are called generic properties. For instance, "a generic polynomial does not have a root at zero," or "A generic matrix is invertible." As another example, "If $f:{M}\rightarrow {N}$ is a smooth function between smooth manifolds, then a generic point of  $N$ is not a critical value of  $f$." (This is by Sard's theorem.)\\
There are many different notions of "generic" (what is meant by "almost all") in mathematics, with corresponding dual notions of "almost none" (negligible set). The two main classes are:\\
( I ) In measure theory, we say that a property is generic if it holds almost everywhere. This means that a property holds for all points except for a set of measure zero.\\
( II ) In topology and algebraic geometry, a generic property is one that holds on a dense open set, or more generally on a residual set, with the dual concept being a nowhere dense set, or more generally a meagre set.\\
In discrete mathematics, one uses the term "almost all"  to mean cofinite (all but finitely many), cocountable (all but countably many), for sufficiently large numbers, or, sometimes, asymptotically almost surely. The concept is particularly important in the study of random graphs.\\
As mentioned above, in topology and algebraic geometry, a generic property is one that holds on a dense open set, or more generally on a residual set (a countable intersection of dense open sets), with the dual concept being a closed nowhere dense set, or more generally a meagre set (a countable union of nowhere dense closed sets). Same definition is used in the theory of Dynamical Systems also.
However, "denseness" alone is not sufficient to characterize a generic property. This can be seen even in the real numbers, where both the rational numbers and their complement, the irrational numbers, are dense in the set of real numbers. Since it does not make sense to say that both a set and its complement exhibit typical behavior, both the rationals and irrationals cannot be examples of sets large enough to be typical. Consequently we rely on the stronger definition above which implies that the irrationals are typical and the rationals are not. We also note that none of these sets is an open subset of real numbers under usual topology.\\
For applications, if a property holds on a residual set, it may not hold for every point, but perturbing it slightly will generally land one inside the residual set, and these are thus the most important case to address in theorems and algorithms.
For function spaces, a property is generic in $C^r{(M, N)}$ if the set holding this property contains a residual subset in the $C^r$ topology. Here $C^r{(M, N)}$ is the function space whose members are continuous functions with $r$ continuous derivatives from a manifold M to a manifold N. Since this topology is often used in the mathematics literature ( called Whitney-$C^r$ topology ), we describe it below in details.
We also note that the space $C^r{(M, N)}$ of $C^r$ mappings between $M$ and $N$  is a Baire space and hence any residual set is dense. This property of the function space is what makes generic properties typical.\\
In general theory of relativity, there are properties of a space-time which are generic. For example, Hawking [6] remarks that a property of a space-time being stably causal is a generic property. Second example of a generic property is that of linearization stability ( we refer the reader to the review article by Fischer and Marsden [11] and to [12,13] for details on linearization stability and related results ). In the class $\mathcal{V}$ of space-times which admit a compact Cauchy hypersurface of constant mean curvature, the subclass $\mathcal{V}_K$ of space-times which are linearization stable form an open and dense subset of $\mathcal{V}$ under $C^\infty$ - topology. Third example of a generic property is the "generic condition" in the Hawking-Penrose singularity theorems. Beem and Harris [14] proved that the set of vector fields satisfying this condition is open and dense in the set of all vector fields under a suitable topology. Fourth instance is the theorems proved by Ringstrom [15,16] in the quest of proving the strong cosmic censorship conjecture. He proves that under suitably defined topology, the set of initial data evolving into Einstein vacuum equations is open and dense in the set of all initial data. This has been proved for (${T^3}\times{R}$)-Gowdy space-times. Fletcher [7] discusses certain causal properties of a space-time in the light of genericity under a given topology on the space of Lorentz metrics.\\ 
When it comes to gravitational collapse of type I matter fields in spherically symmetric space-times, where we consider regular initial data evolving into gravitational collapse, we have proved (cf.[10]) that the set of initial data which leads the collapse to black holes forms an open subset of the set of all initial data under a suitable function space topology. Same is true for the initial data set which leads the collapse to a naked singularity. Neither of these sets are dense in the parent set. However, we proved that these sets have a non-zero measure under a suitable definition of a measure on an infinite dimensional space. \\
Thus, in Section 2, we discuss stability of different global properties of a space-time under suitable definition of topology on the space $Lor(M)$. We also discuss some properties which are generic under a suitable choice of topology. If we choose a different topology, then some undesirable properties of a space-time become generic (cf. [7]). In Section 3, as a proto-type, we discuss the stability and genericity aspects of the outcome of gravitational collapse of inhomogeneous dust in spherically symmetric space-times with respect to small perturbations in the regular initial data from which the collapse evolves. As mentioned above, in [10], these concepts are generalised to gravitational collapse of type I matter fields.  In Section 4, we discuss properties of space-times mentioned above which are generic in the sense of definition as existence of open and dense subset in the appropriate topological sense. In Section 5, we make concluding remarks on both these important issues of stability and genericity.\\

\section{Stability of global properties of a space-time in general theory of relativity} 

\vspace{10mm}

Each Lorentzian metric defined on a four dimensional manifold satisfying Einstein field equations represents a space-time. It determines the geometry (cone structure ) and causal structure of space-time. Until early 1960's,  researchers working in the General Theory of Relativity used to study only special solutions of Einstein field equations. However, with the development of causal structure theory by Hawking and Penrose, and later by other co-workers  ( cf. Hawking-Ellis [17], Joshi[18,19] ), a new avenue was opened  in the study of space-time structure, which was global in nature. Methods were used from topology and global differential geometry to study global structure of space-time which ultimately led to the proof of singularity theorems [17,18]. Hierarchy of a number of causality conditions was discovered in these studies which were mainly global properties of a space-time manifold. ( for recent review of these results, see Minguzzi [20] and  Chrusciel [21]). The important question that then arose was about establishing stability of these  properties in a suitable topological sense. For this it became necessary to define an appropriate topology on the space $Lor(M)$ of Lorentz metrics on a given space-time manifold.

Thus, Hawking [6], Lerner [22] and Beem and Ehrlich [23,24]  defined a Whitney $\ C^{r} $ - topology and other topologies on $Lor(M)$  and studied stability of global properties of space-time.

As we shall see below, definition of stability of a property requires the concept of an open subset of the space of Lorentz metrics. This in turn depends on the topology chosen on the space of such metrics. In this section we first discuss different ways of defining topology on $Lor(M)$ and then discuss global properties which are stable or unstable under some of these topologies. 

Let M be a $\ C^{\infty} $ Hausdorff paracompact four dimensional manifold. We further assume that M is non-compact without boundary. Let Lor(M) denote the space of $\ C^{r} $ Lorentz metrics on M and Con(M) denote the quotient space formed by identifying all pointwise globally conformal metrics $\ g_{1} = \Omega  g_{2} $ on M where $\ \Omega: M \rightarrow (0,\infty) $ is a smooth function. Let $\ \pi : Lor(M) \rightarrow Con(M) $ denote the natural projection map which takes each Lorentzian metric g on M to the set $\ \pi(g) = [g]$ of all Lorentzian metrics on M which are pointwise globally conformal to g. Given $\ [g] \in Con(M)$ , set $\ C$\ (M,g) $ = \pi^{(-1)}([g]) \subseteq Lor(M) $. In general relativity, we say that a curvature condition or a causality condition for a space-time $\ (M,g) $ is $\ C^{r} $ - stable in $L $ or $(M)$ ( respectively $C$ on $(M)$ ) if the validity of the condition for $\ (M,g) $ implies the validity of the condition for all $\ g_{1} $ in a $\ C^{r} $ - open neighbourhood of $g$ in $L$or$(M)$ (respectively $C$on$(M)$ ). More generally, a stable condition for a set of metrics is one which holds on an open subset of such metrics.

In order that the  notion of \emph{open set} is well-defined, we need a topology on $L$or$(M)$. This can be done in more than one way, as follows :\\
To define a topology on $L$or$(M)$, we need the
concept of a \emph{distance} between two Lorentz metrics on $\ M $. For
this, following S. W.Hawking [6], we can consider a positive
definite metric e on $\ M $ (this can always be done, since $\
M $ is paracompact). This metric can then be used to define
covariant derivatives of tensor fields on $\ M $ and also to
measure the magnitude of such tensor fields and their derivatives.
With this, we can define how near together the derivatives of two
metrics are at each point of $\ M $. We have to consider different
possibilities. \\
1.  The metrics can be required to be near only on compact regions
of the manifold and the behaviour near infinity is unrestricted.
This means, if $\ g $ is a Lorentz metric, \emph{U} a compact set
of $\ M $ and $\ \epsilon_{i} (0 \leq i \leq r)$ a set of
continuous positive functions on $\ M $, the neighbourhood $\ B
(U, \epsilon_{i}, g) $ of g can be defined as the set of all
Lorentz metrics whose $\ i ^{th}$ derivatives $\ (0 \leq i \leq r)
$ differ from those of g by less than $\ \epsilon_{i}$ on
\emph{U}. The set of all such $\ B (U, \epsilon_{i}, g) $ for all
\emph{U}, $\ \epsilon_{i}$ and $g$ form a sub-basis for the
\emph{$\ C^{r} $ compact-open topology} for Lor(M). In
other words, the open sets in this topology are unions and finite
intersections of the
$\ B (U, \epsilon_{i}, g) $.\\
2.  The requirement that the sets \emph{U} should be compact can
be removed and \emph{U} can be taken to be $\ M $. This means that
nearby metrics must be nearby everywhere and must have the same
limiting behaviour at infinity. This topology is called
 \emph{open topology} for $L$ or $(M)$.\\
3.  Define the set $\ F (U, \epsilon_{i}, g) $ as the set of all
metrics whose $\ i^{th}$ derivatives differ from those of $\ g $
by less than $\ \epsilon_{i}$ and which coincide with g outside
the compact set \emph{U}. The neighbourhood $\ B ( \epsilon_{i},
g)$ is then defined as the union of $\ F (U, \epsilon_{i}, g) $
for all compact sets \emph{U}. The neighbourhoods $\ B (
\epsilon_{i}, g)$ form a sub-basis for the  fine topology on
$L$or$(M)$. This topology is finer than the open topology,
which in turn is finer than the compact open topology. This means
that there are more open sets in the fine topology than in the
open topology and still more than in the compact open topology.

Classical $\ C^{r} $ derivatives can be replaced by weak /
generalized derivatives and we can obtain Sobolev $\ W^{r}
$-topologies by demanding that instead of requiring the difference
between the derivatives to order r of two nearby metrics to be
small (point-wise), we require the integrals of squares of the
differences of weak derivatives to order r to be small. The
squares and the integrals are here defined with respect to
positive definite metric e on $\ M $. Thus a $\ C^{r} $-tensor
field is also a $\ W^{r} $ field, and by using Sobolev embedding
theorem it follows that a $\ W^{s} $ tensor field for $\ s >
\frac{n}{2}+r = \frac{4}{2}+ r = r+2 $ say $\ s = r + 3 $, is a $\
C^{r} $- field. This means that a $\ W^{r+3} $ topology is finer
than the corresponding $\ C^{r} $- topology which in turn, is
finer than the $\ W^{r} $ - topology ( bigger norm - topology has
more number of open sets than smaller norm - topology ). $\ W^{r}
$- topologies play an important role in the Cauchy problem in
General Relativity ( cf Hawking and Ellis [17] ).

The topologies discussed above do not use the specific properties
which emerge from the symmetry and signature of a Lorentzian
metric.

The topology respecting the causal structure was defined by
Bombelli, Lee, Meyer and Sorkin [25,26] as follows:

They use the fact that causal structure and conformal structure
are the same when the Lorentzian metrics are future and past-
distinguishing. They define a set of functions which compare the
volume elements and causal structure of two metrics $\ g $ and $\
 \tilde{g}$ separately. When restricted to $\ C^{2} $- future and
past-distinguishing metrics, this topology becomes Hausdorff. This
means that for every point $\ p$, there exists a unique maximal
geodesic with starting point $\ p$ and initial direction $\ X_{p}
\in T_{p}M$. This geodesic depends continuously on p and $\ X_{p}
$ i.e. $\ \forall \ (p, X_{p}) \in TM $ and for every open
neighbourhood $\ N $ of $\ exp \  (p, X_{p})$  in $\ M $, there
exists an open neighbourhood \emph{U} in TM such that $\ exp \ (q,
X_{q})\in  N , \forall \ (q, X_{q}) \in U $  whenever $\ exp \  (q,
X_{q}) $ is defined. This definition also satisfies the condition
that for any diffeomorphism $\ \psi , d (g, \tilde{g}) =
d(\psi^{*} g, \psi ^{*} \tilde{g}).$ \\
The definition of topology given by Bombelli and Meyer  [25] is as follows: \\
One can find pseudo-distances which distinguish between metrics
with different volume elements at $\ p \in M $. Using such a
distance between local volume elements, they construct a distance
between metrics in a given conformal class by

$\ d^{M}(g,\tilde{g}) = \displaystyle\sup_{x,y \in M} \mid log
\frac{\sqrt{-g(x)}}{\sqrt{-\tilde{g}(x)}}\mid $

Then, they define a family of non-local functions, characterizing
the fact that causal structures of two arbitrary metrics  $\ g $ and $\ \tilde{g} $
agree down to the volume scale $\ \lambda $, by

$\ d^{C}_{\lambda}(g,\tilde{g}) = \displaystyle\sup_{x,y \in M ,
V(A)
> \lambda} \frac{V(A \Delta A^{'})}{ V (A \cup A^{'})}$ \\
where $\ A $ is the Alexandrov neighbourhood defined by $\ x $ and
$\ y $ in the metric $\ g $ and   $\ A^{'}$ is the corresponding
neighbourhood in the metric $\ \tilde{g} $. Here V (R) denotes the
volume of a region $\ R \subset M $ in the metric $\ g $ and $\ A
\Delta A^{'} $ denotes the symmetric difference $\ (A \sim A^{'})
\cup (A^{'} \sim A)$ . We note that for a fixed value of $\
\lambda , d^{C}_{\lambda}  $ is not symmetric, nor does it satisfy
the triangle inequality. However, the set of functions $\ \{ d^{v}
; d^{C}_{\lambda}; \lambda \in ( 0, \infty ) \}$  induces a
uniform structure on Lor(M).

More recently, Noldus [27] has defined a \emph{new topology} on Lor(M). Noldus has generalized
the above definition of \emph{metric} function $\ d^{C}_{\lambda}(g,
\tilde{g})$ and discussed uniform structure in details.

Here, the main issue is to construct a fully
diffeomorphism - invariant metric topology. Noldus modifies the
work of Bombelli et al and tries to solve this problem \emph{as
much as possible}. His method
is functional analytic. Two important concepts needed are:\\
(i) the choice of topology of the diffeomorphism group (Schwartz
topology) and \\
(ii) the concept of amenability.\\

As remarked by Noldus, the topology on Lor(M)/Diff(M) constructed in [26] is unique, however there exist many pseudo-distances which might generate this uniform topology. It is not known whether the topology on Lor(M)/G as constructed by Noldus is uniquely determined or not. It depends on the generating pseudo-distance of the uniform topology on Lor(M). However, when M is compact with boundary, one can also take the quotient Lor(M)/Diff(M).
However, Noldus topology has much better continuity properties with respect to
group actions. Another thing is, one can also raise the question whether the topology on Lor(M) is locally arcwise connected or not. Diff(M) is locally arcwise connected in the $ \mathcal{FD}$ topology (Refer [28] for definition) so we have that for $\ \phi $ sufficiently small and $\ g \in L $ or $(M)$ that $\ \phi^{\ast}g $ is arcwise connected to g by a path in Lor(M) which corresponds to a path in Diff(M) from the identity to $\ \phi $.

Finally, though Noldus topology is a generalization of topology given in [26], there is no specific direct relationship between this topology and topologies defined earlier on $L$ or $(M)$. It would be interesting to investigate such relationships. However, in some special cases, say for dimension 2, Noldus topology is the same as Euclidean topology.

Thus, so far, we have discussed results regarding the
topologies that can be defined on $L$or$(M)$. Which of these
topologies should be used in a given situation, depends on the
properties one wishes to consider. \\

We now consider stability and instability of some global properties of a space-time with respect to some of the topologies discussed above. We shall also see how stability of a certain property can change to instability if one considers a different topology.

Consider the following causal properties :\\
\textbf{Stable causality }: We say that the stable causality condition holds on a space-time M if the space-time metric ( Lorentz metric ) $\ g $ has an open neighbourhood in the $\ C^{0} $ open topology such that there are no closed timelike curves in any metric belonging to the neighbourhood.\\

Using $\ C^{0}$ topology here will not make any difference, but we could  not use compact-open topology because in this topology each neighbourhood of any metric contains closed timelike curves. Thus, stable causality condition means that one can expand the light cones slightly at every point without introducing closed timelike curves. Fletcher [7] has recently proved some results related to compact-open topologies which illustrate that these topologies are not appropriate for the definition of stable causality. In particular, Fletcher proves that chronology violating space-times form a dense subset of $L$or$(M)$ in any $C^k$ compact open topology. As a consequence of this, no Lorentz metric is stably causal in any $C^k$ compact-open topology. Not only this, Fletcher further proves that chronology violating space-times are generic in $L$or$(M)$ in any $C^k$ compact-open topology. Since a causal theory of space-time demands naturally that any physically reasonable space-time should satisfy the chronology condition, $C^k$ compact-open topologies are not suitable for discussion of stable causality. Fletcher [7] has rightly remarked that the open topologies defined by Lerner are too fine to treat convergence and continuity, whereas compact-open topologies, which are naturally defined through geometric continuity, are too coarse for stable causality.  Since stability of a property is defined through the existence of an open set, it is easier for a property to be stable in a finer topology because there are more open sets available. Thus, if a property is stable on a certain set in a given topology $T$, it is stable in every topology finer than $T$. On the other hand, the concept of convergence of a sequence depends on some aspects of 'every' open neighbourhood of its possible limit. Thus, it is easier for a sequence to converge in a coarser topology, since there are less open sets which should fulfill the given condition. This means, if a sequence converges in a given topology $T$, then it converges in every topology coarser than $T$. These facts show that concepts of stability and convergence are in contrast to each other. Thus, some topologies could be fine enough for the stability of some properties, but too fine for certain sequences to converge, like with the open topology, or vice versa, like with compact-open topology. Fletcher concludes with the apt remark that "Instead of trying to decide what the 'right' topology is for all problems, we should let the details of particular problems guide the choice of an appropriate topology". We refer the reader to Fletcher [7] for more detailed discussion and proofs of these issues. \\

\textbf{Global hyperbolicity} : A space-time $ M $ is said to be globally hyperbolic if it is strongly causal and for any two points $ p,q  \in M$, $ \ J^{+}(p) \cap J^{-}(q) $ is compact.
We refer to [17,18] for detailed discussion of this property and its equivalent versions.
As mentioned above, a property defined on $\ (M,g) $, $\ g \in Lor(M) $, is said to be $\ C^{r}$  stable if it holds on a $\ C^{r}$  open subset ( of Lor(M) ) containing $\ g $. Similarly, a property defined on $\ (M,g) $, $\ g \in Lor(M) $, which is invariant under the conformal relation is said to be conformally  stable if it holds for an open set of equivalence classes in the quotient ( or interval ) topology ( cf. Geroch [29] ) on Con(M). The continuity of the projection map $\ \pi $ implies that any conformally stable property defined on Lor(M) is also $\ C^0 $ stable on Lor(M). Moreover, since the fine $\ C^{r}$  topology is strictly finer than the fine $\ C^{s}$  topology on $L$or$(M)$ for $\ r > s $, any conformally stable property defined on $L$ or$(M)$ is also $\ C^{r}$  stable for all $\ r \geq 0$.

Thus we have the following results  (cf. Hawking [6], Geroch [29], Lerner[22], Beem[24]; also see the book [30] by Beem, Ehlrich and Esley for detailed discussion and proofs ) :\\
1.	Stable causality is conformally stable and hence also $\ C^{r}$ stable in Lor(M) for  all $\ r \geq 0 $. \\
2.	Global hyperbolicity is conformally stable and hence also $\ C^{r}$  stable in Lor(M) for  all $\ r \geq 0 $.\\
3.	If $\ (M,g) $ is a Lorentzian manifold such that $\ g(v,v) \leq 0 $ and $\ v \neq 0 $ in TM imply $\ Ric(g)(v,v) > 0 $, then there is a fine $\ C^{2}$  neighbourhood U(g) of g in Lor(M) such that for all $\ g_{1} \in U(g)$, the relations $\ g_{1}(v,v) \leq 0 $ and $\ v \neq 0$ in TM imply $\ Ric(g_{1})(v,v) > 0 $. This shows that energy condition is a stable property.

However, Williams [31] showed that both \emph{geodesic completeness} and \emph{geodesic incompleteness} may fail to be stable. These two properties are $\ C^{0} $ stable for definite spaces, but for all signatures (s,r) with $\ s\geq 1 \ , \ r \geq 1 $, one can construct examples for which these properties are unstable. (See Del Riego and Dodson [32] for the reasons for these instabilities). However, for Robertson - Walker space-times, geodesic incompleteness can be proved to be stable. See, for example [23].

As mentioned above, global hyperbolicity is a stable property in the set of all time-oriented Lorentz metrics on a fixed manifold. However, A.Garcia Parrado and M. Sanchez [33]  have proved that the causal structures of Minkowski and Einstein static space-times remain stable, whereas that of de Sitter space-time become unstable. More precisely, they prove the following theorem :

For any neighbourhood  $\ \mathcal{U}$ in a $\ C^{r} $ -Whitney topology, $\ r = 0, 1, . . . ,\infty $
 of de Sitter space-time, there is a space-time $\ V \in \mathcal{U}$ such that
V is not isocausal to $\ \mathbf{S}^{n}_{1} $ .
Thus, the causal structure of de Sitter space-time $\ \mathbf{S}^{n}_{1} $ is unstable.

Here, isocausality is defined as follows :\\
Let $\ \Phi  : V_{1} \rightarrow  V_{2} $ be a global diffeomorphism between two manifolds.
We say that the Lorentzian manifold $\ V_{2} $ is causally related to $\ V_{1} $ by $\ \Phi $, denoted
$\ V_{1} \prec_{\Phi}  V_{2} $, if for every causal future-directed $\ \overline{u} \in  T(V_{1}),   \Phi_{\ast}\overline{u} \in T(V_{2})$ is causal future directed too. The diffeomorphism $\ \Phi $ is then called a causal mapping. $\ V_{2} $ is said to be
causally related to $\ V_{1} $, denoted simply by $\ V_{1} \prec  V_{2} $, if there exists a causal mapping $\ \Phi $
such that $\ V_{1} \prec_{\Phi}  V_{2} $. Also, two Lorentzian manifolds $\ V_{1} $ and  $\ V_{2} $ are called causally equivalent
or isocausal if $\ V_{1} \prec  V_{2} $ and $\ V_{2} \prec  V_{1} $. The relation of causal equivalence is denoted by
$\ V_{1} \sim V_{2} $.

In simple language,  isocausal space-times have the same causal structure.
The authors also prove that there are infinitely many different globally hyperbolic causal structures, and thus different conformal ones on $\ R^{2} $. Another interesting result in this paper is that  plane wave solutions with the same number of positive eigenvalues in the frequency matrix have the same causal structure. Thus these solutions have equal causal extensions and causal boundaries.

In a very recent work by Navarro and Minguzzi [34], the authors consider Geroch interval topology and 
prove that global hyperbolicity is stable in the space-time metrics. They prove that every globally hyperbolic space-time admits a Cauchy hypersurface which remains Cauchy under small perturbations of the space-time metric. Moreover they prove that if the space-time admits a complete timelike Killing field, then the light cones can be widened preserving both global hyperbolicity and the Killing property of the field. These results about global hyperbolicity are not contradictory because topologies are different.  In fact, Geroch's interval topology is one of the coarsest topologies that can be given on the space of (conformal classes of) metrics, and hence the stability in this topology is particularly strong. There is only one other important topology that has been discussed above and which is coarser than Geroch's interval topology: the compact-open topology [6].

In this topology, the metric light cones are bounded only inside a compact set of space-time. Thus a property such as global hyperbolicity  need not
be stable in this topology.

Specifically, we note that in all these works, the researchers have considered the concept of stability referring to a particular topology on the space of space-time metrics. Thus these results concern stability of a property of a space-time considered as a whole. They do not consider the problem of stability of a property under evolution.

Thus, in all these works, no evolution equation such as Einstein's field equations, is imposed on the space-time. Such considerations  require a different kind of study like  \emph{Linearized stability} [11 - 13] or stability of end states of collapse [8-10] or proving global existence of solutions of Einstein field equations involving collisionless matter ( cf. Rendall [35] ). We touch first two aspects below.

Finally, we remark that it would be interesting to study stability of space-time properties with respect to the topology developed by Noldus, in the sense that which of the properties are stable and which are not. \\

\section{Stability and Genericity of initial data set in gravitational collapse}

This section is based on the papers [8,9,10] and Saraykar and Joshi [36].
 
 As the theory of gravitational collapse in general relativity
evolved over a period of past four decades, it is now well-known that
dynamical gravitational collapse of a massive matter cloud can end
in either a black hole or a naked singularity final state, for
spherical spacetimes with a variety of matter fields and also in
many non-spherical models. Gravitational collapse has been studied
by many authors in detail (see for example [18] and references
therein). Existence of black hole (BH) or naked singularity (NS)
as endstates of collapse is obtained depending on the choice of
initial data from which the collapse evolves, as was shown by
Dwivedi and Joshi [37,38].

A natural question which then arises is, whether these occurrences
and the collapse outcomes in terms of BH or NS final states are
stable and generic with respect to the regular initial data on an
initial spacelike surface from which the gravitational collapse
develops.

From such a perspective, in the case of inhomogeneous dust,
Saraykar and Ghate [8] showed that the occurrence of NS and BH is
stable with respect to small variations in initial data functions
in the sense that initial data set leading the collapse to NS (or
BH) forms an open subset of the full initial data set under a
suitably defined topology on the space of initial data. The
authors assumed there the definition of genericity as given in the
theory of dynamical systems (see e.g. [39]), and it was then argued
that the NS occurrence is stable but not generic as per that 
definition. It is to be noted, however, that the occurrence 
of black holes also then turns out to be non-generic according 
to such a criterion. Therefore, the definition of genericity, 
as used in the dynamical systems studies would not be adequate 
to be used for discussing
the gravitational collapse outcomes. An important point here is,
work of different researchers over a period of last two decades
has shown that the class of initial data set which leads to NS is
disjoint and `fully separated' from that which leads to BH. This
is true for radiation collapse, inhomogeneous dust collapse and
also for general type I matter fields ([8,10, 18, 40,41]). \\
To examine this point, we consider here the situation of the
inhomogeneous dust collapse following [8]: The spacetime metric
in the case of inhomogeneous dust is given by
\begin{equation}
    ds^2=-dt^2+\frac{{R'}^2}{1+f(r)}dr^2+R^2d\Omega^2 \; ,
\end{equation}
and the energy momentum tensor and field equations are,
\begin{equation}
    T^{ij} = \rho{\delta^i}_t{\delta^j}_t \; ,
\end{equation}
\begin{equation}
   \rho(t,r) = \frac{F'}{R^2R'}
\end{equation}
\begin{equation}
    \dot{R}^2 = {\frac{F(r)}{R}} + f(r)
\end{equation}
where $T^{ij}$ is the energy-momentum tensor, $\rho$ is the total
energy density and $F(r)$ and $f(r)$ are arbitrary functions of
$r$. The dot denotes a derivative with respect to time, while a
prime denotes a derivative with respect to $r$. Integration of
equation $(4)$ gives
\begin{equation}
 t - t_{0}(R) = \frac{-R^{\frac{3}{2}}G
(\frac{-fR}{F})}{\sqrt{F}}
\end{equation}
where the function $G(x)$ takes value  2/3 at $x = 0$, and
is expressed as inverse sine and inverse hyperbolic sine for other
ranges of $x$ (see e.g. [3,4] for exact expressions for $G(x)$.
Following the root equation method of [10], the condition
for existence of a naked singularity or black hole is given in
terms of the function $\Theta_u(r)$ described as follows ([11]):
\begin{equation}
\Theta_{u}(r) =
\frac{1}{\sqrt{g}(3r^{2}g+r^{3}g^{'})}[\frac{(\frac{Q^{'}}{Q}-
\frac{g^{'}}{g})}{\sqrt{1+\frac{Q}{G}}}
+ (\frac{g^{'}}{g}-\frac{3Q^{'}}{2Q})G(\frac{-Q}{g})]
\end{equation}
where the mass function $F(r)$ and energy function $f(r)$ are
written as
\begin{equation}
F(r) = r^3g(r)
\end{equation}
and
\begin{equation}
f(r) = r^2Q(r)
\end{equation}
Moreover $g(r)$ and $Q(r)$ are sufficiently smooth 
(at least $C^1$)
functions satisfying regularity and energy conditions. 
It is assumed that $g(r)$ satisfies
(i) $g(r) > 0$ and (ii) $rg'(r)+ 3g > 0$. These are 
positivity of mass and energy conditions.

Then, if the value of the function $\Theta_{u}(r)$ at $r = 0$ 
is greater than $\alpha$ where,
\begin{equation}
\alpha = \frac{13}{3} + \frac{5}{2} \times \sqrt{3},
\end{equation}
then the tangent to a nonspacelike curve will be positive, {\it
i.e.} future directed nonspacelike curves will reach the
singularity in the past. In other words, singularity will be
naked, not covered by an event horizon. If this condition is
reversed, we get a black hole.

We now consider \emph{A} to be the class of all continuous 
functions $A(r)$ defined on $[0,r_b]$. Consider a subclass of 
\emph{A},
denoted by $\emph{A}_1$, consisting of functions $A(r)$, such
that $ A(0) > \alpha$. Now consider the equation
\begin{equation}
\Theta_u(r) = \emph{A(r)},
\end{equation}
regarded as differential equation in
$Q(r)$) where $\ \Theta_u(r) $ is given by (6).

Then the following result was proved in [8] :
Given a function $g(r)$ satisfying the above conditions, there are infinitely
many choices of function $A(r)$ in the class $\ \emph{A}_1 $ such that
for each such choice of
$A(r)$, there exists a unique function $Q(r)$ such that the initial
data ${(g(r), Q(r))}$
leads the collapse to a naked singularity. Thus, the conditions
on $g(r)$ and $A(r)$ leading the
collapse to NS are, 

(i) $g(r) > 0$, (ii) $rg'(r)+ 3g > 0$ and (iii) $\ A(0) > \alpha$.\\

Since the existence of $Q(r)$ is guaranteed by $g(r)$ and $A(r)$, we
can consider the set $\emph{N}$
of all ${(g(r), A(r))}$ (instead of ${(g(r), Q(r))}$), satisfying the
above conditions, as the set of initial data leading the collapse to
NS. As proved in [4], this set forms an open subset of
$\ \emph{G}\times\emph{A} $ where $\ \emph{G}$ is the space
of all $C^1$ functions defined on the interval $[0,r_b]$, 
and \emph{A} is as above. We note that since the consideration 
of function $A(r)$ comes from equation (6) which contains 
$g(r)$, $Q(r)$ and their first derivatives, it is sufficient to 
assume that the functions $A(r)$ are continuous. Similarly, the set $\ \emph{B}$ of all ${(g(r), A(r))}$ 
satisfying the conditions,

(i) $g(r) > 0$, (ii) $rg'(r)+ 3g > 0$, and (iii) $\ A(0) < \alpha$,

will lead the collapse to a black hole, and similar arguments
imply that $\ \emph{B}$ is also
an open subset of $\ \emph{G}\times\emph{A}$. In this sense, 
both these BH and NS occurrences in collapse are stable.
It is also clear that $\ \emph{N}$ and $\emph{B}$ are disjoint.
Therefore, none of them can be dense in $\ \emph{G}\times\emph{A} $.
Nevertheless, each of these sets are substantially big and 
it can be shown that they have a non-zero positive 
measure [10].\\

It is thus clear that if we follow strictly the dynamical 
systems definition of genericity, then both the outcomes of 
collapse, namely NS and BH would be non-generic. Thus, it is
reasonable to argue that a change in the definition of genericity
is desirable. This change is also justified by the work of other
relativists who used the nomenclature 'generic' in the sense of
'abundance' or existence of an open set of non-zero measure
consisting of initial data leading the collapse to BH or NS as in
the case of scalar fields or for AdS models (see {\it e.g.} 
[42-44] and references therein, but see also [45] where genericity
is defined in terms of codimension). We note that in general 
it is clear from the definitions of `stability' and `genericity' used
that these concepts depend upon the topologies under consideration. 
Thus a given property may be stable and generic in
some topologies and not so in others. Which of the topologies is
of physical interest will depend upon the nature of the property
under consideration.\\
It follows that the dynamical systems definition of `genericity'
needs to be modified if we desire to have black holes as generic
outcomes of gravitational collapse of dust. Dust collapse is
clearly one of the most fundamental collapse scenarios, as the
classic Oppenheimer-Snyder homogeneous dust collapse model is at
the very foundation of the modern black hole physics and its
astrophysical applications.\\

We could therefore formulate an appropriate and physically 
reasonable criterion of genericity for the dust collapse outcomes 
as follows:\\
We assume that the collapse begins with a regular initial 
data, with weak energy condition and other regularity conditions 
satisfied, {\it e.g.} that there are no shell crossings $R'=0$ 
as the collapse evolves. The initial data, namely $F(r)$ and 
$f(r)$ (or $g(r)$ and $A(r)$) allow for the formation of 
both black holes and naked singularities, and we call each 
of these outcomes to be generic if the following 
conditions are satisfied:

    (i) The set of initial data which evolves the collapse to naked 
singularity (or black hole) is an open subset of the full space of 
initial data.\\
    (ii) If we impose a positive measure on the space of initial data, 
then the set of initial data with each of these outcomes should have 
a non-zero measure in the total space.\\

The first condition above means given an initial data point 
$F_1(r)$ and $f_1(r)$, which evolves the collapse to naked singularity 
(black hole), there should be an open neighborhood of this data 
point such that each initial data in this neighborhood also 
evolves to the same outcome. The second condition here means that 
each of these outcomes are substantially big in the full 
space of initial data.\\

We see from the consideration above that this holds true for 
dust collapse. In other words, an outcome of collapse, either in 
terms of a black hole or naked singularity is called `generic' 
if there exists a subset or region of the initial data space 
that leads the collapse to such an outcome, and which has 
a positive measure. Then actually how `big' such a region or
the subspace of the initial data would be, depends on the 
collapse model being considered. For example, for the Vaidya 
radiation collapse, each of the regions going to BH or NS 
seem to be both finitely big and with a non-zero measure, and 
for dust collapse also they seem to have essentially 
equal 'sizes'.\\

Thus, we observe the following: With the above definition 
of genericity, we find that both the outcomes of dust collapse, 
namely black holes and naked singularities are generic 
and also stable (see also [10] for a recent discussion on 
perfect fluid collapse). Since denseness depends upon the 
parent set chosen as well as the choice of topology, choosing 
the definition such as above looks physically reasonable. 
In other words, it is reasonable to argue that a change in 
the definition of genericity which will make both these
collapse final states generic is desirable.\\

\section{More Generic properties occurring in general theory of relativity} 

Maintaining the definition of genericity in terms of existence of an open and dense subset, as in the theory of Dynamical Systems, we now discuss certain properties which are known to be generic in General Relativity :

1. Stable causality is a generic property 
 The region in $L$ or $(M)$ in which stable causality holds lies in the interior of the region on which ordinary causality holds. Also the region in which ordinary causality is violated, is open in $C^0$ open topology. Hence the union of this region with the region on which stable causality holds is an open dense set in $L$or$(M)$. It is thus generic for a metric either to violate causality or to be stably causal.
Conjecture : ( Hawking ) : Stably causal metrics are dense in the causal metrics.

2. Linearization Stability is a generic property :\\
We begin with the following definition :\\
Definition : Let $\Phi: X \longrightarrow Y$ be a non-linear differential operator between Banach spaces or Banach manifolds of maps $X$ and $Y$ (over a compact manifold ) . Consider the equation $\Phi(x)$ = $y_{0}$ for $y_{0} \in  Y$.\\
Let $ T_x{X} $ denote the tangent space to $ X $ at $x \in X$, and let \\ $D\Phi(x)$ : $T_x{X} \longrightarrow T_y{Y}$, with $y$ = $\Phi(x)$, be the Frechet derivative of $\Phi(x)$ at $x$. Thus to each solution $x_0$  of $D\Phi(x)$ = $y_0$ ,  $D\Phi(x_0).h = 0$, $h \in T_{x_{0}}X$, is the associated system of linearized equations about $x_0$, and a solution $h \in T_{x_0}X$ of linearized equations is an infinitesimal deformation ( or first order deformation ) of the solution $x_0$.
If for each solution $h$ of linearized equations , there exists a curve $x_t$   of exact solutions of $D\Phi(x) = y_0$  which is tangent to $h$ at $x_0$  i.e. $x(0) = x_0$  and $[(d / dt)(x_t )]_{|_{t = 0}} = h$, then we say that equation $\Phi(x) = y_{0}$ is linearization stable at $x_0$, and deformation $h$ is called integrable.\\
Useful criterion to prove linearization stability is as follows :\\
Theorem: Let $X$ and $Y$ be Banach manifolds, and $\Phi: X \longrightarrow Y$ be a $C^1$-map. Let $x_0 \in X$ be a solution of $\Phi(x) = y_{0}$. Suppose $D\Phi(x_0)$ is surjective with splitting kernel . Then the equation $\Phi(x)$ = $y_{0}$ is linearization stable at $x_0$.\\
( Splitting kernel means:  $T_{x}X = Range (D\Phi)^*(x) + Ker D\Phi(x)$). \\
Proof uses the Implicit Function Theorem.\\
We now discuss Linearization stability of Einstein's equations in case where Cauchy (spacelike) hypersurface is a compact 3- manifold without boundary. 
In analogy with above definition, we define linearization stability of Einstein field equations :\\
We write Einstein equations for vacuum space-time as : $Ein(^4{g}) = 0$, where Ein denotes Einstein tensor. Let $^4{g}_0$ be a solution of Einstein equation. Then linearized equation is given by $DEin(^4{g}_0).^4{h}  = 0$. If for every $^4{h}$ satisfying linearized Einstein equation, there is a curve ($^4{g}_ {t}$) of exact solutions such that $[^4{g}_t]_{|t = 0}  = ^4{g_0}$ , and $[(d / dt)(^4{g_t} )]_{|_{t = 0}} =  ^4h$, then $^4{g}_0$ is called linearized stable.  In this case $^4{ h}$  is called integrable.
Not every $^4{ h}$ satisfying Linearized equation is integrable:
Brill and Deser [46] considered space-time $({T^3})\times R$ and showed that there are solutions of linearized equation which are not integrable.
For more details, we refer the reader to the review article by Fischer and Marsden [47].\\
Important results about Linearization Stability are as follows:\\
(1) For space-time admitting a compact constant mean curvature (CMC) space-like hypersurface $(D\Phi)^*(g,\pi)$ is elliptic. Here $g$ is the induced Riemannian 3-metric on specelike hypersurface, say $M$, and $\pi$ is the conjugate momenta corresponding to $g$ in the standard ADM formalism.\\
(2) Let space-time $(V, ^4{g})$ be fixed. The space of Killing fields of $^4{g}$  is isomorphic to the kernel of $(D\Phi)^*(g,\pi)$. Here, we note that an elliptic operator has a finite dimensional kernel.\\
(3) If $(V, {^4{g}})$  has no Killing fields, then it is linearization stable.\\
Thus, if $(V, ^4{g})$ is linearization stable , ker $(D\Phi)^*(g,\pi)$ is trivial.\\
Thus, space-times admitting compact CMC hypersurfaces can admit only finite number of independent Killing fields. It will not be out of place to mention that similar results hold for Einstein field equations coupled with matter fields such as scalar fields, electro-magnetic fields and Yang-Mills fields. See, for example, $[47,48,49]$.
For full analysis of the structure of the space of solutions of Einstein's field equations in the presence of Killing fields, we refer the reader to Fischer, Marsden and Moncrief $[12]$ and Arms, Marsden and Moncrief $[13]$.(See also Saraykar $[50]$).\\
We now consider the class $\mathcal V$ of all space-times ( equivalently class of Lorentz metrics ) possessing compact Cauchy hypersurfaces of constant mean curvature, and  use results of Beig, Chrusciel and Schoen [51] and P. Mounoud [52] to argue that the subclass ${\mathcal V}_K$  of $\mathcal V$ possessing no Killing fields forms an open and dense subset of $\mathcal V$.
Combining this result with above result for linearization stability, we conclude that the class of space-times possessing compact Cauchy hypersurfaces of constant mean curvature which are linearization stable forms an open and dense subset of $\mathcal V$. In this sense, property of linearization stability is generic.
The argument in support of this goes as follows :
Beig, Chrusciel and Schoen [51] have proved that the class of space-times with compact Cauchy hypersurface of constant mean curvature and which possesses no Killing initial data (KID), forms an open and dense subset of the class of space-times with compact Cauchy hypersurface of constant mean curvature. Due to covariance property of Einstein field equations, Killing property of initial data will be carried throughout the evolution. Thus this property is possessed by space-time as a whole. In other words, KIDs are in one-to-one correspondence with Killing vectors in the associated space-time.
They also prove that above result remains true if Cauchy surface is asymptotically flat or asymptotically hyperbolic.
More generally, Mounoud [52] proved the following theorem :
Theorem : Let $V$ be a compact manifold with dimension $\geq 2$ and $M_{p,q}$ be the set of smooth pseudo-Riemannian metrics of signature $(p,q)$ on $V$. Then the set $G =\left\{ g \in  M_{p,q}: Is(g) = Id \right\} $ contains a subset that is open and dense in $M_{p,q}$  for the $C^\infty$-topology. Here, $Is(g)$ denotes isometry group of $g$. 
For Riemannian metrics, similar result was well-known as Ebin [53] proved in 1970 that the set of Riemannian metrics without isometries on a compact manifold is open and dense. Similar but more general results were noted and proved by Henrique de A. Gomes [54] by using Ebin-Palais slice theorem.
Finally, in the context of general relativity, it is interesting to note that it is not yet fully known if the above generic result is valid when constant mean curvature condition is removed.\\

3. Generic condition is generic : \\

Let $(V,g)$ denote a space-time and $R_{ab}$  denote Ricci tensor corresponding to the Lorentz metric $g$. Let $\gamma(v)$ denote a timelike or null curve in $V$ and let $p = \gamma(v_1)$. Let $K$ denote a general tangent vector. Then the following result holds (Cf. Hawking-Ellis [17], Prop. 4.4.5, page 101) :\\

Result 1 : If $R_{ab}K^{a} K^{b} \geq 0$ holds everywhere and if at $p = \gamma(v_1)$, $K^{c} K^{d} K_{[a}R_{b]cd[e}K_{f]}$ is non-zero, then there will be $v_0$ and $v_2$ such that $q=\gamma(v_0)$ and $r=\gamma(v_2)$ will be conjugate along $\gamma(v)$ provided $\gamma(v)$ can be extended to these values.\\
 
Hawking-Ellis explain that it is reasonable to assume that in a physically realistic space-time, every timelike or null geodesic will contain a point at which the quantity $K^{c} K^{d} K_{[a}R_{b]cd[e}K_{f]}$ is non-zero. This condition is called 'generic condition'. It can be satisfied by a single tangent vector, or by every tangent vector to a null or timelike curve. In the latter case, we say that a space-time itself satisfies the generic condition. It is well-known [17,18] that the generic condition for a space-time plays an important role in the proof of singularity theorems. Beem and Harris[14] prove that this Generic condition is generic in the sense that at a given point of spacetime, if we consider the tangent space at this point, then the set of tangent vectors satisfying the generic condition is open and dense in the tangent space. Beem and Harris prove a series of results which are algebraic in nature, and then the above result follows as a consequence of these. We describe these results in brief :\\
A tangent vector $K$ is called non-generic if it is non-zero and if $K^{c} K^{d} K_{[a}R_{b]cd[e}K_{f]}$ = $0$. Thus, a causal geodesic satisfies the generic condition if and only if its velocity vector is not everywhere non-generic. \\
Following series of results have been proved in {14] : \\
Result 2 : The set of non-generic vectors, together with the zero vector, is a closed set. Furthermore, any null vector which is a limit of non-generic non-null vectors or of strongly non-generic null vectors is itself strongly non-generic. \\
Hence the set of generic vectors is an open set.\\
Result 3 : The entire vector spave $V$ is non-generic if and only if it is flat.\\
Result 4 : Let $V$ be four dimensional. Suppose $V$ has $5$ non-null non-generic vectors, with $4$ forming a basis and the fifth not in the plane spanned by any two of those basis vectors. Then $V$ is flat.\\
Result 5 : Let $W$ be a subspace of $V$ with codimension of $W$ = $1$. If $W$ is non-generic, then $V$ is flat.\\
Result 6 : Let $W$ be a subspace of $V$ with dim($W$)= $m$. Suppose $W$ has the following non-null non-generic vectors $\left\{X_i : {1 \leq i \leq m}\right\}$, which span $W$; and for each $i<j$, $Y_{ij}$ in span $({X_i, X_j})$, such that $(X_i, X_j,Y_{ij})$ is in general position for span $({X_i, X_j})$. Then $W$ is non-generic.\\
Applying this result to dim($V$) = $4$, and dim ($W$) = $3$, we get the following corollary of result $6$ :\\
Result 7 : If $V$ has a subspace $W$ of codimension 1, so that $W$ contains a set of non-generic vectors which is open in $W$, then $V$ is flat.\\
Denseness property of generic vectors now follows from this corollary (Result 7) as argued below :\\
If the generic vectors do not form a dense set, then there is an open set of non-generic vectors, whose intersection with any subspace is open. This implies, by above corollary, that $V$ is flat; which gives a contradiction. Thus, generic vectors form a dense set. That it forms an open set, follows from Result 2 mentioned above. Thus, set of generic vectors froms an open and dense subset of the set of all tangent vectors at a given point, i.e.the tangent space at that point.\\
This then proves that generic condition applied to a tangent vector is really a generic property.\\
 
Our last example of genericity is from the work of Ringstrom where he proves strong cosmic censorship conjecture for a certain class of space-times.\\
4. Set of initial data (Cauchy surface ) from which a space-time develops, in the sense of maximally globally hyperbolic development, is open and dense in the class of all initial data in a suitable topological sense	( Ringstrom [15,16] ) : \\

Einstein's vacuum equations can be viewed as an initial value problem, and given initial data there is one part of space-time, the so-called maximal globally hyperbolic development (MGHD), which is uniquely determined up to isometry. However, it is sometimes possible to extend the space-time beyond the MGHD in inequivalent ways. Hence, the initial data do not uniquely determine the space-time, and in this sense the theory is not deterministic. Here, it is then natural to make the strong cosmic censorship conjecture, which states that for generic initial data, the MGHD is inextendible. Since it is unrealistic to hope to prove this conjecture in all generality, it is natural to make the same conjecture within a class of space-times satisfying some symmetry condition. Ringstrom, in a series of two papers, proved strong cosmic censorship in the class of $T^ 3$-Gowdy spacetimes.
In the first paper, the author focuses on the concept of asymptotic velocity. Under the symmetry assumptions, Einstein's equations reduce to a wave map equation with a constraint. The range of the wave map is the hyperbolic plane. The author introduces a natural concept of kinetic and potential energy density. The important result of this paper is that
the limit of the potential energy as one lets time tend to the singularity for a fixed spatial point is $0$ and that the limit exists for the kinetic energy.
In the second paper [16], the author proves that the set of initial data $G_i$ is open with respect to the $\ C^1$ topology and dense with respect to the $\ C^\infty$ topology, such that the corresponding space-times have the following properties:

First, the MGHD is $C^2$-inextendible. Second, following a causal geodesic in a given time direction, it is either complete, or a curvature invariant, the Kretschmann scalar is unbounded along it (in fact the Kretschmann scalar is unbounded along any causal curve that ends on the singularity). \\
 
 \section{Conclusion}

As we are aware, the concepts such as `stability' and `genericity'
which are so important for physical considerations, are not
well-defined in the Einstein gravitation theory. Also, there are many aspects of stability on which researchers have spent their time and efforts. For example, by considering odd and even parity perturbations of a given spherically symmetric space-time metric, like Schwarzschild or Reissner-Nordstrom, people have studied stability of these solutions. For this, we refer the reader to an excellent book by S. Chandrasekhar [55]. Depending upon the context, and problem under study, researchers have studied stability of different phenomena from different angles.  \\
Thus, definition of 'Stability' seems to depend upon the topology we choose and class of perturbations we deal with. Consider, for example, the Schwarzschild space-time. It is globally hyperbolic. But if we throw a smallest charge into it, then the resulting space-time is Reissner-Nordstrom, which is not globally hyperbolic. Moreover it admits a Cauchy horizon, and Cauchy surfaces are lost. As another example, consider the Oppenheimer-Snyder collapse, i.e. the spherically symmetric dust collapse, which is globally hyperbolic. But with a slightest perturbation in density, e.g. the density higher at the center, we loose global hyperbolicity, and the singularity becomes naked. Now choosing a slightly higher density at center corresponds to a certain perturbation of the metric. Thus, it will be interesting to consider such a class, which is of course physically interesting, and  try and see why global hyperbolicity is violated ? So should we call it stable or not ? In the same example, if density is inhomogeneous and if we add slight pressure in the collapsing system, then again the outcome can change from black hole to a naked singularity or vice-versa. Thus the outcome can be called unstable with respect to addition of pressure. Such problems have been studied by Joshi and Malafarina ( see [9] for a detailed review).\\
In addition to this, as we have said and discussed in earlier sections, when the definition involves openness of the data, then the choice of topology plays an important role in deciding stability as well as genericity properties. Thus, we conclude that the definition of stability under consideration depends upon the context and phenomena under study, as well as the choice of mathematical objects like function spaces and topology involved in the definitions. Thus, we can not expect uniqueness in the choice of definition of stability. Just as this is true for problems in general theory of relativity, it is also true in other branches of applied mathematics. \\

Acknowlegdement : The author wishes to express sincere thanks to Prof. Pankaj S. Joshi, Tata Institute of Fundamental Research, Mumbai, for many helpful comments, suggestions and discussions about this work.\\
\vspace{10mm}

 \section{References}
 
  1. Y. Hagihara, Stability in Celestial Mechanics, Kasai Publications, Tokyo, (1957).\\  
  2. I. Prigogine, "From Being to Becoming"', W.H.Freeman and Company, San Francisco, (1980).\\
  3. V. Szebehely, Review of concepts of stability, Celestial Mechanics, 34,49-64(1984).\\
  4. A. Loria and E. Penteley, Stability, told by its developers, In : A. Loria, F. Lamnabhi-Lagarrigue, E. Penteley (Editors)
  : Advanced Topics in Control Systems Theory, Lecture Notes in Control and Information Sciences, Vol.328, 199-258, Springer, London (2006) Chapter 6.\\
  5. R.I. Leine, The historical development of classical stability concepts : Lagrange, Poisson and Lyapunov stability, Nonlinear Dynamics, 59, 173-182 (2010).\\
  6. S.W. Hawking, Gen. Rel. Grav. 1, 393 (1970).\\
  7. S.C. Fletcher, Physicality and Topological Stability in Relativity Theory, (2013).\\
  8. R.V. Saraykar and S.H.Ghate, Class. Quantum Grav. 16, 281-289 (1999).\\
  9. P.S. Joshi and D. Malafarina, Int. J. Mod. Phys. D, 20 2641 (2011), arXiv:1201.3660v1 [gr-qc].\\
  10. P.S. Joshi, D. Malafarina and R.V. Saraykar, Inter. Jour. Mod. Phys. D, Vol.21, No.8, (2012) 1250066 (38 pages).\\
  11. A.E. Fischer A. and J.E. Marsden, Topics in the dynamics of general relativity, in �Isolated gravitating systems in general relativity�, Ed. J. Ehlers,            Italian Physical Society, (1979), 322-395.\\
  12. A.E. Fischer, J.E. Marsden and V. Moncrief, The structure of the space of solutions of Einstein�s equations.I. One Kiling field,Annales de la Institut H.          Poincare, Section A, Vol.33 (2) (1980), 147-194.\\
  13. J. Arms, J.E. Marsden and V. Moncrief, The structure of the space of solutions of Einstein�s equationsII. Several Kiling fields and the Einstein-Yang-Mills        Equations, Annals of Physics, Vol. 144 (1) (1982), 81-106.\\
  14. J.K. Beem and S.G. Harris, Gen. Rel. and Grav. 25(9),939-961(1993).\\
  15. H. Ringstrom, Commun. in Pure and Appl. Math.59, 977-1041,(2006).\\
  16. H. Ringstrom, Ann. of Math., 170, 1181-1240, (2009).\\
  17. S.W. Hawking and G.F.R. Ellis, The large scale structure of space-time, Cambridge University Press, (1973).\\
  18. P.S. Joshi, Global aspects in gravitation and cosmology, Oxford Press, 1993.\\
  19. P.S. Joshi, Gravitational collapse and spacetime singularities, Cambridge press, 2007.\\
  20. E. Minguzzi, The causal ladder and the strength of K-causality, I and II, Class. Quant. Grav. Vol. 25: 015009, 015010 (2008).\\
  21. Piotr T. Chrusciel, Elements of causality theory, arXiv: gr-qc / 1110.6706v1, (2011).\\
  22. D. Lerner, Topology on the space of Lorentz metrics, Commu. Math. Phys., Vol.32 , 19 - 38, (1973).\\
  23. J.K. Beem and P.L. Ehrlich, Geodesic completeness and stability, Math. Proc. of Camb. Phil. Soc., Vol.102, no.2, 319-328, ( 1987).\\
  24. J.K. Beem, Stability of geodesic incompleteness, in Differential Geometry and mathematical Physics, Ed. J.K.Beem and K.L.Duggal, Amer.Math.Soc. (1994).\\
  25. L. Bombelli, J. Lee, D.A. Meyer, R.D. Sorkin, Space-time As a Causal Set, Phys. Rev. Lett. 59, 521-524, (1987).\\
  26. L. Bombelli and D.A. Meyer, The Origin of Lorentzian Geometry, Phys. Lett. A 1414, 226-228, (1989).\\
  27. J. Noldus, A new topology on the space of Lorentzian metrics on a fixed manifold, Class. Quant. Grav. Vol.19, 6075-6107, (2002).\\
  28. A. Kriegl, P. W. Michor,  The convenient Setting of Global Analysis, American Mathematical Society Math. Surv. and Mon. 53, (1997).\\  
  29. R. Geroch, Domain of dependence, Jour.Math.Phys. Vol.11, 437-449, (1970).\\
  30. J.K. Beem, P.E. Ehrlich and K.L. Easley, Global Lorentzion Geometry, Monographs textbooks Pure Appl Mathematics, Dekker Inc., New York, (1996).\\
  31.	P.M. Williams, Completeness and its stability on manifolds with connection, Ph.D.Thesis, Dept. of Maths., University of Lancaster, (1984).\\
  32.	L. Del Riego and C.T.J. Dodson, Sprays, universality and stability, Math. Proc. Camb. Phil. Soc.  Vol.103, 515-534, (1987).\\
  33.	A. Garcia-Parrado and M. Sanchez,  arXiv: math-ph / 0507014 v2.\\
  34.	J.J.B. Navarro and E. Minguzzi, Jour.Math.Phys., Vol. 52, 112504, (2011).\\
  35.  A. Rendall, Theorems on Existence and Global Dynamics for the Einstein Equations, Living Rev. Relativity, 8, (2005), 6,\
      http://www.livingreviews.org/lrr-2005-6. (Update of lrr-2002-6)\\
  36. Ravindra V. Saraykar and Pankaj S. Joshi, A Note on Genericity and Stability of Black Holes and Naked Singularities in Dust Collapse, Clifford Analysis, Cliffors Algebras and Their Applications (CACAA) - www.cliffordanalysis.com, Vol. 3, No. 4, pp. 295-299,(2014), copy right CSP - Cambridge, UK, I and S - Florida, USA, 2014.\\
  37. I.H. Dwivedi and P.S. Joshi, Class.Quant.Grav. 14 (1997) 1223. \\
  38. P.S. Joshi and I.H. Dwivedi, Class.Quant.Grav. 16 (1999) 41-59.\\
  39. R. Abraham and J.E. Marsden, Foundations of Mechanics, 2nd edition, Addison-Wesley (1978)..\\
  40. S.B. Sarwe and R.V. Saraykar, Pramana, a Journal of Physics, 65, 17 (2005).\\
  41. I.H. Dwivedi and P.S. Joshi, Phys. Rev. D 47, 5357 (1993).\\
  42. C. Gundlach, Phys. Rept. 376, 339 (2003).\\
  43. C. Gundlach and J. M. Martin-Garcia, \\ arXiv:0711.4620v1 [gr-qc];\\
  44. T. Hertog, G.T. Horowitz and K. Maeda, Phys.Rev.Lett.92,131101 (2004).\\
  45. D. Christodoulou, Ann. of Math. 149, 183 (1999).\\
  46. D. Brill and S. Deser,  Commu. Math. Phys. 32, 291-304 (1973) \\
  47. R.V. Saraykar and N.E. Joshi, Jour. Math. Phys. 22, 343-347(1981); doi: 10.1063/1.524885.\\
  48. J. Arms, Jour. Math. Phys. 18, 830-833(1977).\\
  49. J. Arms, Jour. Math. Phys. 20, 443-453(1979).\\
  50. R.V. Saraykar, Pramana, a Journal of Physics, vol.20,no.4, 293 - 303 (1983).\\
  51. Robert Beig, Piotr T. Chrusciel and Richard Schoen, KIDs are non-generic, Ann. Inst. H. Poincare, Vol. 6, 155-194 (2005) \\
  52. Pierre Mounoud, Metrics without isometries are generic,  arXiv:1403.0182v1 [math.DG] \\
  53. D.G. Ebin, The manifold of Riemannian metrics, Proc. Symp. Pure Maths. Vol. XVI, (1970), "On the space of Riemannian metrics", Ph.D. Thesis at MIT , 1967.\\
  54. Henrique de A. Gomes, A Note on the Topology of a Generic Subspace of Riem,  arXiv:0909.2208v1 [math-ph]\\
  55. S. Chandrasekhar, The Mathematical theory of black holes, Clarendon Press, Oxford, 1982. \\
   
\end{document}